\documentclass[aps,prd,amsmath,amssymb]{revtex4}

\usepackage{graphicx}
\usepackage{dcolumn}
\usepackage{bm}
\usepackage{amsmath}
\usepackage{epstopdf}
\usepackage{amsfonts}
\usepackage{amssymb}
\usepackage{epsfig}
\usepackage{tabularx}
\usepackage{color}
\usepackage{soul}
\usepackage{multirow}
\usepackage{xcolor}

\usepackage[colorlinks=true,citecolor=blue,urlcolor=magenta,breaklinks]{hyperref}
\newcommand{\be}{\begin{equation}}
\newcommand{\en}{\end{equation}}
\newcommand{\bea}{\begin{eqnarray}}
\newcommand{\ena}{\end{eqnarray}}

\newcommand{\bff}{\begin{figure}}
\newcommand{\eff}{\end{figure}}

\begin{document}

\title{Radial Oscillations of the HESS J1731-347 Compact Object via the Karmarkar Condition in Gravity}

\author{Grigoris Panotopoulos}

\address{
Departamento de Ciencias F{\'i}sicas, Universidad de la Frontera, Casilla 54-D, 4811186 Temuco, Chile.
\\
\href{mailto:grigorios.panotopoulos@ufrontera.cl}{\nolinkurl{grigorios.panotopoulos@ufrontera.cl}} 
}

\begin{abstract}
We model the light HESS J1731-347 compact object (of known stellar mass and radius) within Einstein's General Relativity imposing the Karmarkar condition in gravity for anisotropic stars. The three free parameters of the analytic solution are determined imposing the matching conditions at the surface of the star for objects of known stellar mass and radius. Finally, using well established criteria it is shown that the solution is compatible with all requirements for well behaved and realistic solutions. Furthermore, we study the radial oscillation modes, and we compare to the ones corresponding to an isotropic star modeled by the Tolman IV exact analytic solution obtained long time ago. A comparison between the large frequency separations is made as well.
\end{abstract}

\maketitle

\section{Introduction}

Thanks to Einstein's General Relativity (GR) that was formulated last century \cite{Einstein:1915ca}, we understand and describe a wide range of cosmological and astrophysical aspects. Einstein's gravity is worldwide considered to be both beautiful and successful, since many of its remarkable predictions have been confirmed. One can mention for instance the classical tests discussed in standard textbooks \cite{Weinberg, MTW} as well as more recent ones after the historical LIGO’s first direct detection of gravitational waves emitted from black holes binaries \cite{Ligo1, Ligo2, Ligo3}. For a recent review on tests of GR see e.g. \cite{Asmodelle:2017sxn}.

\vskip 0.15cm

In spite of the mathematical elegance of GR, in most of the cases of interest the study and analysis of realistic physical situations comprises a formidable task, due to the fact that Einstein's field equations are highly non-linear coupled partial differential equations. For that reason, the principle of superposition, which is valid in the theory of linear differential equations and which facilitates the task of obtaining the solution, does not apply in GR. Therefore, finding exact analytic solutions to the field equations of Einstein's gravity has always been an interesting and challenging topic, keeping researchers busy for decades. For exact known solutions to Einstein's field equations, see \cite{ExactSol}.

\vskip 0.15cm

When studying stars the authors usually focus their attention on astrophysical objects made of matter viewed as a perfect fluid, where there is a unique pressure along all spatial dimensions, $p_r=p_\theta=p_\phi$. However, celestial bodies are not necessarily made of isotropic fluids. In fact, under certain conditions matter content of compact objects may become anisotropic. The review article of Ruderman \cite{aniso1} long time ago mentioned such a possibility: there the author makes for the first time the observation that in very dense media anisotropies can be generated by interactions between relativistic particles. In the following years the study on anisotropic relativistic stars received a boost by the subsequent work of \cite{aniso2}. Indeed, anisotropies may arise in many different physical situations, such as phase transitions \cite{aniso3}, pion condensation \cite{aniso4}, or in presence of type 3A super-fluid \cite{aniso5}, to mention just a few.

\vskip 0.15cm

Thanks to the work of \cite{Ovalle:2017fgl}, there is now a new and elegant method which permits us to construct anisotropic solutions starting from an already known seed isotropic solution. The so-called Minimal Geometric Deformation approach was originally introduced in \cite{Ovalle:2007bn} in the context of brane models \cite{RS1,RS2}, it has been extensively applied over the years, and it has been proven to be a powerful tool in the study of self-gravitating objects, such as relativistic stars or black holes \cite{Estrada:2018zbh, Morales:2018urp, Estrada:2018vrl, Ovalle:2018umz}, see also \cite{Ovalle:2008se, Ovalle:2010zc, Casadio:2012pu, Casadio:2012rf, Ovalle:2014uwa}. The aforementioned method was later extended in \cite{Ovalle:2018gic}, and it was applied in \cite{Fernandes-Silva:2019fez}, demonstrating its power and potential. 

\vskip 0.15cm

What is more, the author of \cite{herrera} recently introduced the so called complexity factor, which is a measure of complexity of self-gravitating systems. Technically speaking, the complexity factor appears in the orthogonal splitting of the Riemann tensor. It is easy to see that it vanishes for isotropic spherical configurations or for fluid spheres characterized by homogeneous energy densities, but it may also vanish when the two terms containing density inhomogeneity and anisotropic pressure cancel each other. For articles on stars made of anisotropic matter within the complexity factor formalism see e.g. \cite{comp1, comp2, comp3, comp4, comp5, comp6, comp7, comp8, comp9, comp10, comp11}.

\vskip 0.15cm

Another approach that allows us to find exact analytic interior solutions of relativistic stars is imposing the Karmarkar condition \cite{karmarkar}. The method works equally well for compact objects made of isotropic and anisotropic matter. The strategy of the approach is the following: If one of the metric potentials is assumed, the Karmarkar condition allows us to obtain the other metric potential. After that the energy density and the pressures of the fluid may be computed one by one making use of the field equations. It is worth mentioning that within this approach no equation-of-state (EoS) for the matter content is assumed. For an incomplete list of works obtaining interior solutions via the Karmarkar condition in gravity see e.g. \cite{kar1, kar2, kar3, kar4, kar5, kar6, kar7, kar8, kar9, kar10, kar11}.

\vskip 0.15cm

As far as the inner structure of relativistic compact stars and the underlying EoS are concerned, the most massive pulsars \cite{Demorest,Antoniadis,recent} that have been observed over the last 15 years or so are putting constraints on different EoSs, since any mass-to-radius relationship that predicts a highest mass lower than the observed ones must be ruled out. In order to model a certain astronomical object, it would be advantageous to know both the stellar mass and radius. This is not always the case, since measuring the radius is much more difficult than measuring the stellar mass. As of today there are some good strange quark star candidates, see e.g. Table 5 of \cite{Weber} or Table 1 of \cite{Maxim}, and also the recently discovered massive pulsar PSR J0740+6620 \cite{pulsar1,pulsar2,pulsar3} and the strangely light HESS J1731-347 compact object \cite{hess}, where both the stellar mass and radius are known observationally. Based on modelling of the X-ray spectrum and a robust distance estimate from Gaia observations, the authors of \cite{hess} estimated the mass and the radius of the object to be $M=0.77_{-0.17}^{+0.20} M_{\odot}, R=10.40_{-0.78}^{+0.86} km$. Their estimate implies that this object is either the lightest neutron star known, or a ‘strange star’ with a more exotic equation of state. The recent observation of an exceptionally low mass compact object inside the supernovae remnant HESS J1731-347 challenges the standard neutron star formation models, and it has attracted a lot of attention among the nuclear physics community. Since it is one of the lightest and smallest astronomical objects ever observed, it raises many questions regarding its nature, and opens up the window for different possibilities to explain such a measurement \cite{violeta}.

\vskip 0.15cm

Helioseismology and Asteroseismology in general, a research area that studies the oscillation modes of pulsating stars, is a powerful tool and a widely used technique to probe the inner structure of stars. The precise values of the frequency modes are very sensitive to the underlying physics and the corresponding equation-of-state of the dense matter content of compact objects. Therefore, studying the oscillations and computing the frequencies of the modes we can learn more about the composition of compact objects, see e.g. \cite{Brillante:2014lwa,Kokkotas:2000up,Miniutti:2002bh,Panotopoulos:2017eig,Passamonti:2005cz,Passamonti:2004je,Savonije:2007ay,VasquezFlores:2017tkp,VasquezFlores:2010eq} and references therein. Regarding excitation and detectability, numerous astrophysical mechanisms may excite oscillation modes of stars, such as tidal effects in binaries, starquakes caused by cracks during supernova explosions, magnetic reconfiguration, or any other form of instability \cite{Ex1,Ex2,Ex3,Ex4}.

\vskip 0.15cm

The Kepler and CoRoT missions have already measured the oscillation spectra of solar-like, white dwarf, and red giant stars \cite{corot1,corot2,corot3,Kepler}. Current wave detectors, such as LIGO, cannot detect the radial oscillations studied here due to their low sensitivity at the kHz frequency range. However, the third-generation ground-based gravitational wave detectors are expected to have a much higher sensitivity (by an order of magnitude), such as the Cosmic Explorer \cite{CE} and the Einstein Telescope \cite{ET}. These detections could give us information related to neutron star masses, frequencies, tidal Love numbers, amplitudes of the modes, damping times, and moments of inertia. Upon comparison between accumulated experimental and observational data and well-accepted theoretical predictions, we should be able to probe and infer the EoS of dense matter. 

\vskip 0.15cm

Any exact analytic solution obtained within one of the aforementioned approaches must be physically relevant. In other words it must be finite at the origin, well behaved and realistic, fulfilling all the well-established criteria discussed in the literature. Even better, it should be capable of modeling one or more of the observed stars of known mass and radius. The goal of the present study is twofold. First we propose to model the HESS compact object assuming it is made of anisotropic matter via the Karmarkar condition in gravity. Moreover we shall show that the solution is well-behaved capable of describing realistic astrophysical configurations. Furthermore, we shall study its radial oscillation modes, and we shall make a comparison to the modes of its isotropic counterpart modeled via the Tolman IV solution \cite{Ovalle:2013xla,tolman}.

\vskip 0.15cm

In the present article our work is organized as follows. After this introduction, in the next section we briefly review relativistic compact objects assuming a non-vanishing factor of anisotropy, and we obtain an exact analytic solution. In section 3 we summarize the usual criteria that ensure the existence of realistic solutions, we model the HESS compact object via the solution obtained in section 2, and we demonstrate that it is capable of describing realistic astrophysical configurations. In the fourth section we study the radial oscillation modes of the HESS compact object. For the sake of comparison, we also study radial oscillations of the isotropic counterpart.  Finally, we conclude our work in the last section. Throughout the manuscript we adopt the mostly positive metric signature, $\{ -, +, +, + \}$, as well as geometrical units, in which both the speed of light in vacuum and Newton's constant are set to unity, $c = 1 = G$.

\section{Anisotropic relativistic stars in Einstein's Gravity}

\subsection{Structure equations for fluid spheres}

When seeking interior solutions ($0 \leq r \leq R$, $R$ being the radius of the star), the most general form of a static, spherically symmetric geometry in Schwarzschild-like coordinates, $\{ t, r, \theta, \phi \}$, is given by
\begin{equation}
    d s^2 = - e^{\nu} d t^2 + e^{\lambda} d r^2 + r^2 (d \theta^2 + \sin^2 \theta d \phi^2) 
\end{equation}
with $\nu(r), \lambda(r)$ being two independent functions of the radial coordinate. In the following we shall be working with the mass function, $m(r)$, which is defined by
\begin{equation}
    \displaystyle e^{-\lambda} = 1 - \frac{2 m}{r}.
\end{equation}

In order to find solutions describing hydrostatic equilibrium of compact objects we have to integrate the Tolman-Oppenheimer-Volkoff equations \cite{tolman, OV}
\begin{eqnarray}
    m'(r) & = & 4 \pi r^2 \rho (r)   \\ 
    p_r'(r) & = & - [ \rho(r) + p_r(r) ] \; \frac{\nu' (r)}{2} + \frac{2 \Delta}{r} \\
    \nu' (r) & = & 2 \: \frac{m(r) + 4 \pi r^3 p_r(r)}{ r^2 \left( 1 - 2 m(r) / r \right) } 
\end{eqnarray}
where us usual a prime denotes differentiation with respect to $r$. 

Next, anisotropic matter is described by a diagonal energy-momentum tensor of the form
\begin{equation}
T_a^b = Diag(-\rho, p_r, p_t, p_t)
\end{equation}
where $p_r,p_t$ are the radial and tangential pressure of matter, respectively, $\rho$ is the energy density of the fluid, while the anisotropic factor is defined by
\begin{equation}
 \Delta = p_t - p_r,
\end{equation}
while in the case of stars made of isotropic matter there is a unique pressure in both directions, $p_r=p_t$, and therefore $\Delta=0$.

The equations (3) - (5) are to be integrated imposing at the center of the star appropriate conditions 
\begin{equation}
    m(0) = 0, \; \; \; \; \; p_r(0) = p_t(0) = p_{c},
\end{equation}
where $p_{c}$ is the common central pressure, and therefore the anisotropic factor vanishes at the center of the star, $\Delta(0)=0$. In addition, the following matching conditions must be satisfied at the surface of the object
\begin{equation}
    p_r(R) = 0, \; \; \; \; m(R) = M, \; \; \; \; e^{\nu(R)}=1-2 \frac{M}{R},
\end{equation}
with $M$ being the stellar mass, taking into account that the exterior vacuum solution ($r > R$) is given by the Schwarzschild geometry \cite{Schwarzschild:1916uq}
\begin{equation}
    d s^2 = - (1-2M/r) d t^2 + (1-2M/r)^{-1} d r^2 + r^2 (d \theta^2 + \sin^2 \theta d \: \phi^2) .
\end{equation}

Finally, upon numerical integration, once the mass function and the radial pressure are known, the metric potential $\nu(r)$ is computed by
\begin{equation}
    \displaystyle \nu (r) = \ln \left( 1 - \frac{2 M}{R} \right) + 2 \int^r_R dx \: \frac{m(x) + 4 \pi x^3 p_r(x)}{ x^2 \left( 1 - 2 m(x) / x \right) } .
\end{equation}

\subsection{Anisotropic stars imposing the Karmarkar condition in gravity}

If the metric tensor satisfies the Karmarkar condition \cite{karmarkar}, it can represent an embedding class one spacetime 
\begin{equation}
R_{1414} = \frac{R_{1212} R_{3434} + R_{1224} R_{1334}}{R_{2323}},
\end{equation}
with $R_{2323}$ different than zero. This condition leads to a differential equation given by \cite{kar1}
\begin{equation}
2 \frac{\nu''}{\nu'} +\nu' = \frac{\lambda' e^\lambda}{e^\lambda-1} .
\end{equation}

If the function $\lambda(r)$ is given, upon integration we obtain the relationship between the metric potentials as follows \cite{kar1}
\begin{equation}
e^\nu = \left( C_1 + \frac{1}{C_2} \int dr \sqrt{e^\lambda-1} \right)^2,
\end{equation}
where $C_1$ and $C_2$ are arbitrary constants of integration. Equivalently, if the function $\nu(r)$ is given, the other metric potential upon integration is found to be
\begin{equation}
e^\lambda = 1 + \frac{B^2 (\nu')^2 e^\nu}{4},
\end{equation}
where this time $B$ is an arbitrary constant of integration.

In the discussion to follow we shall assume a metric potential of the form
\begin{equation}
\nu(r) = C + \left( \frac{r}{A} \right)^2,
\end{equation}
which is characterized by two free constant parameters, $A,C$, and which is one of the potentials of the Krori-Barua ansatz \cite{Barua:1975dxq}. It has been used among astronomers in the literature as it generates singularity-free solutions in general relativity \cite{Barua:1975dxq}. Imposing the Karmarkar condition, the second metric potential is found to be
\begin{equation}
e^\lambda = 1 + \frac{B^2 r^2}{A^4} exp(C+(r/A)^2)
\end{equation}
while the mass function is found to be
\begin{equation}
m(r) = \frac{B^2 r^3 exp(C+(r/A)^2)}{2 [A^4+B^2 r^2 exp(C+(r/A)^2)]}
\end{equation}
Next, using the structure equations, the radial pressure and the energy density of the fluid may be computed, and they are given by
\begin{equation}
\rho(r) = \frac{B^2 exp(C+(r/A)^2) [3 A^4 + 2 A^2 r^2 + B^2 r^2 exp(C+(r/A)^2)]}{8 \pi [A^4+B^2 r^2 exp(C+(r/A)^2)]^2}, \; \; \; \; \; \; \rho_c=\rho(0)=\frac{3 B^2 e^C}{8 \pi A^4}
\end{equation}
\begin{equation}
p_r(r) = \frac{2 A^2 - B^2 exp(C+(r/A)^2)}{8 \pi [A^4+B^2 r^2 exp(C+(r/A)^2)]}, \; \; \; \; \; \; p_c=p_r(0)=\frac{2-(B/A)^2 e^C}{8 \pi A^2} ,
\end{equation}
where the central values $p_c,\rho_c$ are found to be finite, while at the same time the radial pressure vanishes when
\begin{equation}
R=A \sqrt{-C+ln\left( \frac{2 A^2}{B^2} \right)},
\end{equation}
which corresponds to the stellar radius. Finally, the anisotropy $\Delta(r)$ is found to be
\begin{equation}
\Delta(r) = \frac{[A^2 - B^2 exp(C+(r/A)^2)]^2 r^2}{8 \pi [A^4+B^2 r^2 exp(C+(r/A)^2)]^2},
\end{equation}
and the tangential pressure, $p_t = \Delta + p_r$, is computed to be
\begin{equation}
p_t(r) = \frac{A^4 [2 A^2 + r^2 - B^2 exp(C+(r/A)^2)]}{8 \pi [A^4+B^2 r^2 exp(C+(r/A)^2)]^2}.
\end{equation}

Finally, the speed of sounds at constant entropy, $c_{s,r}, c_{s,t}$, and the relativistic adiabatic indices, $\Gamma_r, \Gamma_t$, are defined as
\begin{equation}
c_{s,i}^2 \equiv \left( \frac{dp_i}{d \rho} \right)_S =\frac{p_i'(r)}{\rho'(r)} 
\end{equation}
\begin{equation}
\Gamma_{i} \equiv c_{s,i}^2 \left( 1+\frac{\rho}{p_i} \right),
\end{equation}
for both directions, namely radial (r) and tangential (t), $i=r,t$, and they are computed in terms of the known functions $p_r(r), p_t(r), \rho(r)$ and their derivatives.

\section{Modeling the HESS Compact Object}

\subsection{Criteria for Realistic Solutions}

Before we proceed with stellar modeling, we shall report here the requirements for well-behaved solutions capable of describing realistic astrophysical configurations. The well-established criteria discussed in the literature are the following:

\begin{itemize}

\item Causality, i.e. the speed of sound must be lower than the speed of light in vacuum
\begin{equation}
0 \leq c_s^2 \leq 1
\end{equation}

\item Stability based on the relativistic adiabatic index, i.e. its mean value must be larger than a critical value
\begin{equation}
\langle \Gamma \rangle \geq \Gamma_{cr}
\end{equation}
where the critical value is given by \cite{Moustakidis:2016ndw}
\begin{equation}
\Gamma_{cr} = \frac{4}{3} + \frac{19}{21} \: \frac{M}{R}
\end{equation}
while the mean value is computed by \cite{Moustakidis:2016ndw}
\begin{equation}
\langle \Gamma \rangle = \frac{\int_0^R dr \: \Gamma(r) p(r) r^2 e^{(\lambda+3 \nu)/2}}{\int_0^R dr \: p(r) r^2 e^{(\lambda+3 \nu)/2}} .
\end{equation}

\item The energy conditions impose certain constraints on the stress-energy tensor of matter within a given theory of gravity. The acceptable conditions assumed for the energy-momentum tensor are as follows: weak energy condition (WEC), dominant energy condition (DEC), null energy
condition (NEC), and strong energy condition (SEC), see for instance \cite{Ellis, Wald, Frolov}. 
If $\xi_\mu$ and $k_\mu$ are arbitrary time-like and null vectors, respectively, then the conditions for the energy-momentum tensor are expressed with the following inequalities
\begin{equation}      
T^{\mu \nu} \: \xi_\mu \: \xi_\nu \geq 0, \; \; \; \; (WEC)
\end{equation}
\begin{equation}      
T^{\mu \nu} \: \xi_\mu \: \xi_\nu \geq 0 \; \textrm{and} \; T^{\mu \nu} \: \xi_\mu \; \textrm{is a non-spacelike vector,} \; \; \; \; (DEC)
\end{equation}
\begin{equation}      
T^{\mu \nu} \: k_\mu \: k_\nu \geq 0, \; \; \; \; (NEC)
\end{equation}
\begin{equation}      
T^{\mu \nu} \: \xi_\mu \: \xi_\nu - (1/2) T_{\mu}^{\mu} \: \xi^\nu \: \xi_\nu \geq 0 . \; \; \; \; (SEC)
\end{equation}

In particular, in the case of isotropic fluids the energy conditions require that \cite{Panotopoulos:2020uvq, Balart:2023odm}
\begin{equation}      
\rho \geq 0, \; \; \; \; \; \; \rho + 3 p \geq 0, \; \; \; \; \; \; \rho \pm p \geq 0 .
\end{equation}
while in the case of anisotropic fluids the energy conditions take the form
\begin{equation}      
\rho \geq 0, \; \; \; \; \; \; \rho + p_r + 2 p_t \geq 0, \; \; \; \; \; \; \rho \pm p_{i} \geq 0, \; \; \; \; i=r,t.
\end{equation}

\end{itemize}

\subsection{The HESS J1731-347 Object}

\begin{figure*}[ht!]
\centering
\includegraphics[scale=0.95]{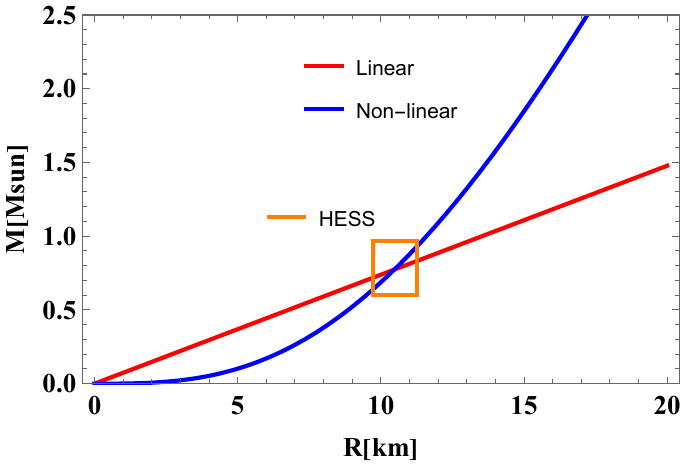} \\
\includegraphics[scale=0.95]{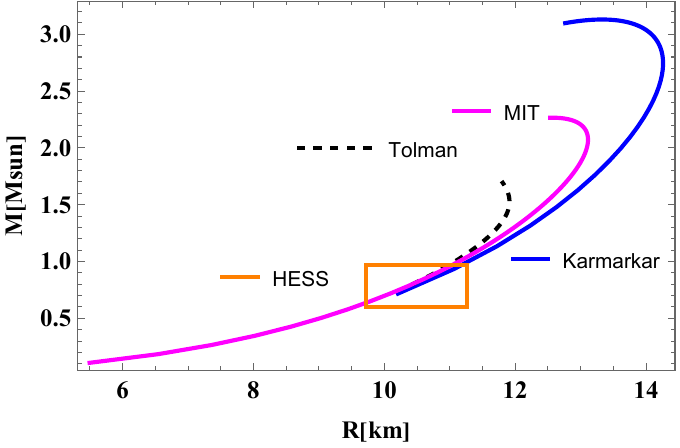} \\
\includegraphics[scale=0.7]{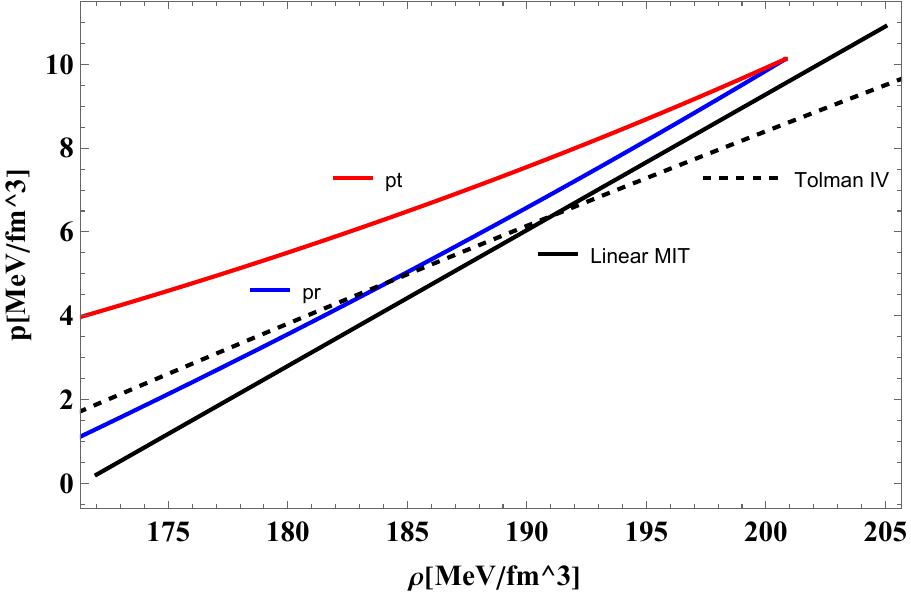} 
\caption{
Mass-to radius-relationships (upper and middle panels), and pressure versus energy density $p-\rho$ (lower panel) for the solution discussed in this work. In the upper panel the analytic expressions for the stellar mass and radius were used, whereas in the middle panel the TOV equations were integrated numerically using the $p-\rho$ relationship. For comparison reasons the M-R relationships corresponding to the MIT bag model as well as the Tolman IV solution are shown as well. The allowed range of the HESS compact object, too, is included. In the lower panel a comparison is made between the radial and tangential pressure versus energy density for the analytic solution discussed in this work, the linear MIT EoS of the model SQSB40, and the $p-\rho$ relationship of the Tolman IV solution.
}
\label{fig:0} 	
\end{figure*}

Imposing the following matching conditions at the surface of the star, $r \rightarrow R$
\begin{equation}
p_r(R)=0, \; \; \; \; \;  m(R)=M, \; \; \; \; \; e^{\nu(R)}=1-2 \frac{M}{R}
\end{equation}
we find the following values of the free parameters
\begin{equation}
A=27.90 km, \; \; \; \; \; \; B=44.64 km, \; \; \; \; \; \; C=-0.39
\end{equation}
considering stellar mass and radius $M=0.77 M_{\odot}$ and $R=10.47 km$, respectively. For those numerical values of $A,B,C$, the central values of the energy density and both pressures are found to be
\begin{equation}
\rho_c = 200.83 \: MeV/fm^3, \; \; \; \; \; \; p_c = 10.13 \: MeV/fm^3.
\end{equation}
Furthermore, the relations between the energy density and both pressures are shown in the lower panel of Fig. \ref{fig:0}. The pressures increase with the energy density, which is the typical behavior of any equation-of-state. Since they seem to be almost linear, we make a comparison with a linear MIT bag model EoS, (extreme model SQSB40) \cite{Gondek-Rosinska:2008zmv}, see the discussion below for more details on that model. The $p-\rho$ relation corresponding to the Tolman IV solution (see subsection 4.2) is also included. What is more, since there is an analytic expression both for the stellar radius and the stellar mass in terms of the constant parameters $A,B,C$, we may generate the M-R relationships displayed in the upper panel of Fig. \ref{fig:0}. The red curve is a straight line corresponding to fixed $B,C$ and varying $A$, whereas the blue curve is a non-linear function that is obtained fixing $A,C$ and varying $B$ or fixing $A,B$ and varying $C$. The two curves meet at the point that represents the HESS compact object modeled in the present work. Finally, using the $p-\rho$ relationship one may solve the TOV equations numerically to obtain the M-R relationship shown in the middle panel of Fig. \ref{fig:0}. It does not come as a surprise that it looks very similar to the M-R profiles of strange quark stars. The $M-R$ profiles of the quark star and of the HESS object modeled via the Tolman solution, too, are shown for comparison reasons. We have included the allowed stellar mass and radius range of the HESS compact object as well.

\smallskip

We comment in passing that although it is not obvious, it turns out that the solution obtained here is unique. It is not difficult to verify this numerically. One of the three matching conditions may be used to express the parameter $C$ in terms of the other two. Then, the other two matching conditions imply certain relations between $A$ and $B$. One may draw the contour plots, and observe that the two curves meet at a single point.

\smallskip

Next, all the quantities of interest, such as sound speed, factor of anisotropy etc., versus $r$ are displayed in the Figures \ref{fig:1} and \ref{fig:2} below. In particular, Fig. \ref{fig:2} shows the speed of sound and the relativistic adiabatic index, in both directions, versus $r$. In the lower panel of Fig. \ref{fig:1} it is observed that the tangential pressure is larger than the radial one, and so the anisotropic factor is positive. For the sake of comparison, we show in the same plot (upper panel of Fig. \ref{fig:1}) the energy density and the pressure of a quark star made of isotropic matter. The red color corresponds to the analytic solution discussed here, whereas the black color corresponds to the quark star assuming a linear EoS of the form (extreme model SQSB40) \cite{Gondek-Rosinska:2008zmv}
\begin{equation}
p_q = k (\rho_q - \rho_0),
\end{equation}
where the parameters $k,\rho_0$ take the numerical values \cite{Gondek-Rosinska:2008zmv}
\begin{equation}
k = 0.324, \; \; \; \; \; \rho_0 = 3.06 \times 10^{17} kg/m^3.
\end{equation}
for $m_s=100 MeV, \alpha_c=0.6$ and $B=40 MeV/fm^3$, with $m_s$ being the mass of the s quark, $\alpha_c$ being the strong coupling constant, and $B$ being the bag constant \cite{Gondek-Rosinska:2008zmv}.
For those values, and for a central pressure $p_{q,c} = 1.275 \: \rho_0$, the mass and radius of the quark star are found to be $M_q=0.79 M_{\odot}, R_q=10.38 km$. We see that the functions $p(r), \rho(r)$ corresponding to the numerical solution lie very close to the ones corresponding to the analytic solution. Therefore, a quark star made of isotropic matter, too, may model the HESS compact object for very similar energy density and pressure.

\smallskip

According to our results, since both energy density and pressures are positive, and $\rho > p_{i}, i=r,t$ all the energy conditions are fulfilled, while at the same time causality is not violated. Moreover, both the pressure and the energy density of the fluid are finite at the center of the star, and they monotonically decrease with the radial coordinate. Finally, since $\Gamma  > \Gamma_{cr}$, we conclude that the stability criterion is met as well.

\smallskip

It is worth noticing that the solution found here is based on the use of the Karmarkar condition and in the particular choice of the metric potential. Following another approach or choosing another metric potential will lead to different expressions for the quantities of interest and different numerical values of the free parameters of the solution. Furthermore, if another approach is adopted (for example one may assume a concrete EoS and a certain relationship between energy density and anisotropic factor or a certain mass function), the new solution does not have to satisfy the Karmarkar condition.

\smallskip

Furthermore, as a conjectural corollary, let us argue an analogy to the proposal by \cite{Drago0,Drago1,Drago2} of the so called two family scenario: Two distinct families of compact objects could coexist in Nature. The first family (hadronic neutron stars) could explain some observations, while the second family (strange quark stars) could explain some others. According to that scenario, objects characterized by small stellar masses and radii are hadronic, whereas objects characterized by large stellar masses and radii are made of quark matter. Given the TOV results shown in the middle panel of Fig. \ref{fig:0}, since according to the blue curve $M_{max} \approx 3 M_{\odot}$, it can be confirmed that the obtained $M-R$ relationship is compatible with the original Drago proposal. It must be emphasized, however, that when anisotropies are present, like in the present study, things can and will change compared to stars made of isotropic matter, since when $\Delta > 0$ the EoS can support a higher maximum stellar mass.

\section{Radial oscillation modes of relativistic stars}

\subsection{Equations for perturbations}

Assuming a system with radial motion only, Einstein's field equations may be used to compute the spectra of a static equilibrium structure. Considering the radial displacement $\Delta r$ and the pressure perturbation $\Delta P$ as small perturbations to the background solution, the dimensionless quantities $\xi \equiv \Delta r/r$ and $\eta \equiv \Delta P/P$ are found to satisfy a first order system of coupled differential equations as follows \cite{pert1, pert2} 
\begin{equation}\label{ksi}
     \xi'(r) = -\frac{1}{r} \Biggl( 3\xi + \frac{\eta}{\Gamma} \Biggr) - \frac{P'(r)}{P+\rho} \xi(r),
\end{equation}
\begin{equation}\label{eta}
    \begin{split}
          \eta'(r) = \xi \Biggl[ \omega^{2} r (1+\rho/P) e^{\lambda - \nu } -\frac{4P'(r)}{P} -8\pi (P+\rho) re^{\lambda} \\
     +  \frac{r(P'(r))^{2}}{P(P+\rho)}\Biggr] + \eta \Biggl[ -\frac{\rho P'(r)}{P(P+\rho)} -4\pi (P+\rho) re^{\lambda}\Biggr] ,
    \end{split}
\end{equation}
with $\omega$ being the frequency oscillation mode. The above equations for perturbations allow us to study the radial oscillation modes of stars made of isotropic matter. In the case of anisotropic stars, there are some additional terms that are proportional to the anisotropic factor, and therefore the linear system of coupled perturbations reads \cite{Arbanil:2021ahh}
\begin{equation}\label{ksi}
     \xi'(r) = -\frac{1}{r} \Biggl( 3\xi + \frac{\eta}{\Gamma} \Biggr) - \left( \frac{P'(r)}{P+\rho} + \frac{2 \Delta}{r P \Gamma} \right) \xi(r),
\end{equation}
\begin{equation}\label{eta}
    \begin{split}
          \eta'(r) = \xi \Biggl[ \omega^{2} r (1+\rho/P) e^{\lambda - \nu } - \frac{4P'(r)}{P} -8\pi (P+\rho) re^{\lambda} \frac{P+\Delta}{P} \\
     +  \frac{8 \Delta}{r P} + \frac{r(P'(r))^{2}}{P(P+\rho)}\Biggr] + \eta \Biggl[ -\frac{\rho P'(r)}{P(P+\rho)} -4\pi (P+\rho) re^{\lambda}\Biggr] ,
    \end{split}
\end{equation}
where now $P,\Gamma$ are the radial quantities, $P(r)=p_r(r), \Gamma(r)=\Gamma_r(r)$.

\smallskip

Those two coupled differential equations, Eqs.~\eqref{ksi} and \eqref{eta}, are supplemented with two boundary conditions: one at the center, where $r=0$, and another at the surface, where \mbox{$r=R$}. The equation Eq.~\eqref{ksi} must be finite at the center, and therefore the boundary condition at the origin requires that
\begin{equation}
    \eta = -3\Gamma \xi 
\end{equation}
must be satisfied. Moreover, the equation Eq.~\eqref{eta} must be finite at the surface and hence
\begin{equation}
    \eta = \xi \Biggl[ -4 +(1-2M/R)^{-1} \Biggl( -\frac{M}{R} -\frac{\omega^{2} R^{3}}{M}\Biggr)\Biggr]
\end{equation}
must be satisfied, where $M$ and $R$ correspond to the mass and radius of the star, respectively. The frequencies are computed by
\begin{equation}
\nu = \frac{\omega}{2\pi} = \frac{s \: \omega_0}{2\pi} ~~(\text{kHz}),
\end{equation}
where $s$ is a dimensionless number, while $\omega_0 \equiv \sqrt{M/R^3}$.

\smallskip

Those equations represent the Sturm-Liouville eigenvalue equations for $\omega$. The solutions provide the discrete eigenvalues $\omega_n^{2}$ and can be ordered as 
\begin{equation*}
\omega_0 ^{2} < \omega_1 ^{2} <... <\omega_n ^{2}, 
\end{equation*}
where $n$ is the number of nodes of the corresponding eigenmode for a star of a given mass and radius.

\smallskip 

We use the shooting method analysis, where one starts the integration for a trial value of $\omega^2$ and a given set of initial values that satisfy the boundary condition at the center. We integrate towards the surface, and the discrete values of $\omega^2$ for which the boundary conditions are satisfied correspond to the eigenfrequencies of the radial perturbations.

\smallskip

Finally, one commonly used indicator is the so called large frequency separation
\begin{equation}
\Delta \nu_n = \nu_{n+1} - \nu_n, \; \; \; n=0,1,2,3,...
\end{equation}
which can be easily computed once the spectrum is known. In other words, it is the difference between consecutive modes, and it can be shown to relate to stellar mass and radius \cite{tassoul, miglio, ilidio}.


\begin{table}
\caption{
Frequencies (in kHz) of radial oscillation modes both for isotropic and anisotropic stars. We have considered $M = 0.8 M_{\odot}$ and $R = 10.42 km$ in the case of the isotropic object, and $M = 0.77 M_{\odot}$ and $R = 10.47 km$ in the case of the anisotropic star.
}
\label{tab:1}       
\begin{tabular}{lll}
\hline\noalign{\smallskip}
Mode order $n$  & Isotropic star & Anisotropic star  \\
\noalign{\smallskip}\hline\noalign{\smallskip}
0  &  4.48 &  5.81 \\ 
1  & 10.27 & 12.62 \\ 
2  & 15.76 & 19.20 \\
3  & 21.17 & 25.72 \\
4  & 26.55 & 32.22 \\ 
5  & 31.93 & 38.72 \\ 
6  & 37.29 & 45.20 \\ 
7  & 42.65 & 51.68 \\
 \noalign{\smallskip}\hline
\end{tabular}
\end{table}


\subsection{Tolman IV solution}

Before we proceed with the discussion of the numerical results regarding the frequencies and corresponding eigenfunctions of the HESS compact object, let us briefly review the Tolman IV solution \cite{tolman}. Among the hundreds of exact analytic solutions to the field equations of GR, the Tolman solution is one of the few solutions with physical meaning, in the sense that there is a well defined stellar mass and radius, subluminous speed of sound, while at the same time the energy density and the pressure of the fluid are positive definite, regular at the origin, and monotonically descreasing functions of the radial coordinate throughout the star \cite{Ovalle:2013xla}.

We may start assuming a metric potential of the form
\begin{equation}
e^{\lambda(r)}  =  \frac{1+2 (r/a)^2 }{[1-(r/r_0)^2] [1+(r/a)^2]}.
\end{equation}
Next, using the definition of the mass function as well the first structure equation, the energy density and the mass function are computed to be
\begin{eqnarray}
\rho(r) & = & \frac{1}{8 \pi r_0^2} \: \frac{6 r^4 + (7 a^2+2 r_0^2) r^2+3 a^4+3 a^2 r_0^2}{(a^2+2 r^2)^2}, \\
m(r) & = & \frac{r^3}{2 r_0^2} \: \frac{r^2+a^2+r_0^2}{a^2+2 r^2}. 
\end{eqnarray}
Finally, making use of the other two structure equations, the pressure of the fluid and the other metric potential, $\nu(r)$, are found to be \cite{Ovalle:2013xla}
\begin{eqnarray}
p(r) & = & \frac{1}{8 \pi r_0^2} \: \frac{r_0^2-a^2-3 r^2}{2 r^2+a^2}, \\
e^{\nu(r)} & = & b^2 [ 1+(r/a)^2 ] .
\end{eqnarray}

Clearly, the solution is characterized by three constant parameters $a,b,r_0$. As a check, it is straightforward to verify that all structure equations are satisfied. We see that both the pressure and the energy density remain finite at the center of the star, while at the same time the pressure becomes zero at a finite radius $R$, which is found to be 
\begin{equation}
R = \frac{r_0}{\sqrt{3}} \: \sqrt{1-\frac{a^2}{r_0^2}}.   
\end{equation}
Finally, the speed of sound, $c_s$, and the relativistic adiabatic index, $\Gamma$, for the Tolman IV solution are found to be
\begin{equation}
c_s^2(r) = \frac{p'(r)}{\rho'(r)} = \frac{2 r^2+a^2}{2 r^2+5 a^2}
\end{equation}
\begin{equation}
\Gamma(r) = c_s^2 \left( 1+\frac{\rho}{p} \right) = 2 \frac{(r^2+a^2) (2 r_0^2+a^2)}{(2 r^2+5 a^2) (r_0^2-a^2-3 r^2)}.
\end{equation}
Recently, it was shown in \cite{Panotopoulos:2024imo} that the Tolman IV solution is capable of modeling the HESS compact object, provided that the latter is isotropic, if the parameters of the solution take the numerical values
\begin{equation}
a=25.07 km, \; \; \; \; \; r_0=30.89 km, \; \; \; \; \; b=0.81
\end{equation}
leading to a stellar mass of $M=0.80 M_{\odot}$ and a stellar radius of $R=10.42 km$.

\subsection{Discussion of the results}

\begin{figure*}[ht!]
\centering
\includegraphics[scale=0.8]{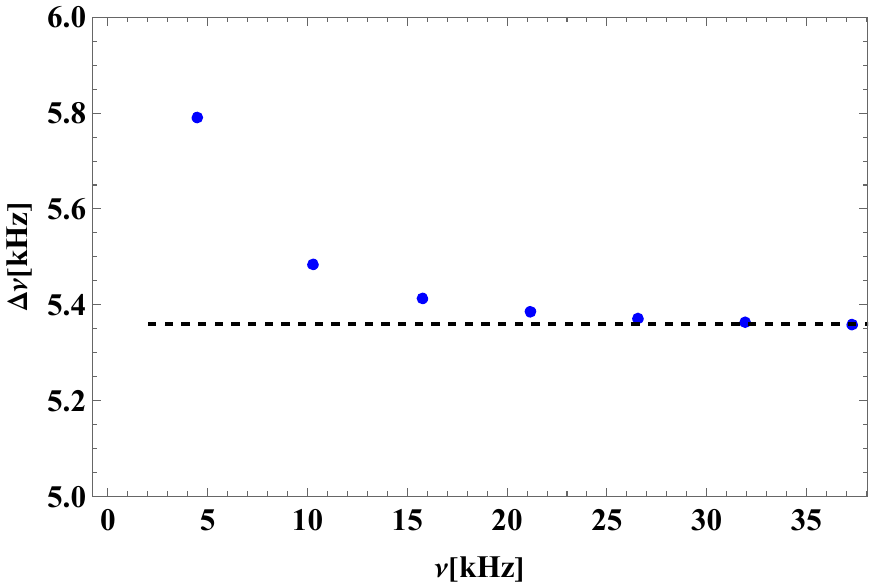} \\
\includegraphics[scale=0.8]{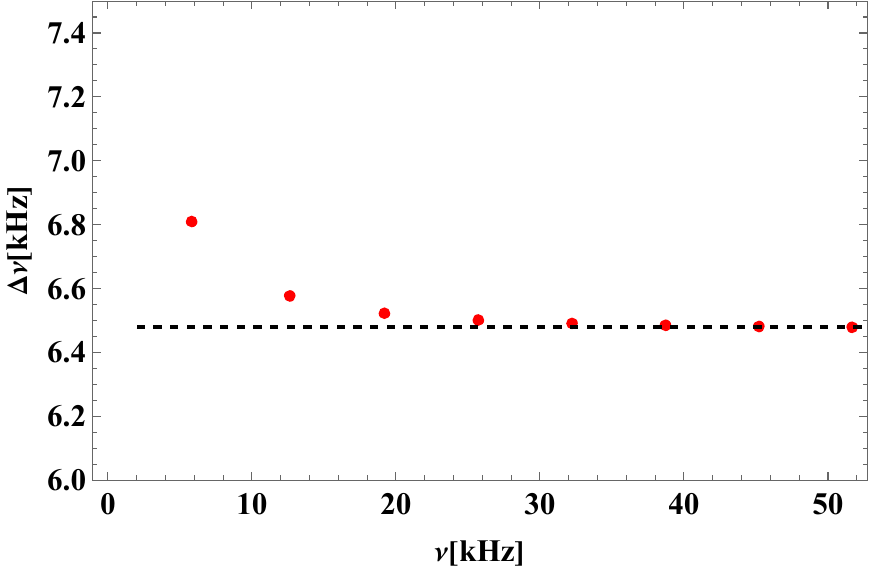} 
\caption{
Large frequency separations for isotropic star (Tolman IV solution, upper panel) and anisotropic object (via Karmarkar condition, lower panel) versus frequency for the HESS compact object. At higher excited modes the separation tends to a constant value, $\Delta \nu^{iso} \approx 5.36 \: kHz$ in the case of isotropic star, and $\Delta \nu^{aniso} \approx 6.48 \: kHz$ in the case of anisotropic star.
}
\label{fig:3} 	
\end{figure*}

\begin{figure*}[ht!]
\centering
\includegraphics[scale=1]{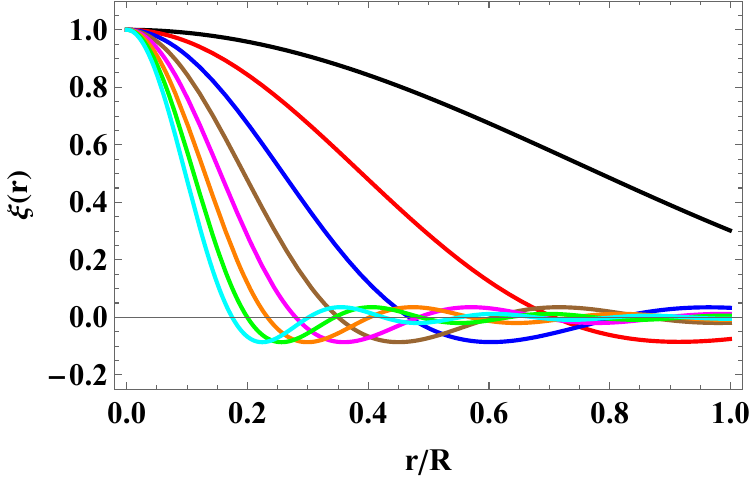} \\
\includegraphics[scale=1]{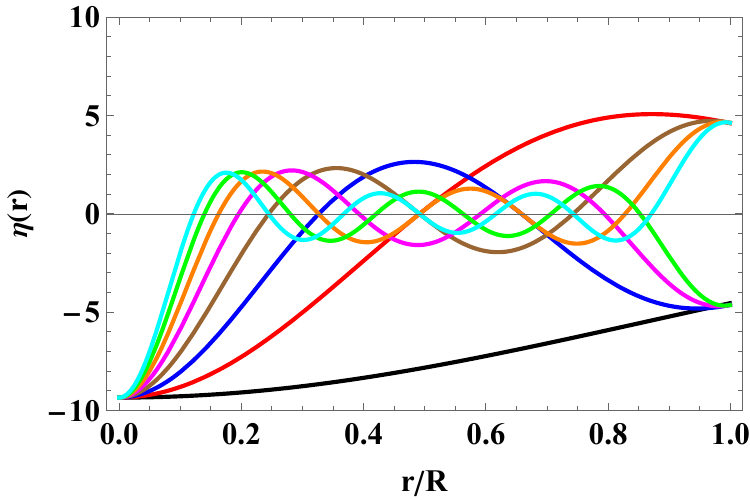}
\caption{
Radial profiles of eigenfunctions versus normalized radial coordinate, $r/R$ for the fundamental mode, $n=0$, as well as seven excited modes, $n=1,2,...7$. The upper panel corresponds to the $\xi(r)$ function, while the lower panel to $\eta(r)$ function.
}
\label{fig:4} 	
\end{figure*}

The frequencies of the 8 lowest modes (fundamental, $n=0$ and seven excited modes, $n=1,2,...,7$) are shown in Table 1 both for the anisotropic HESS compact object (modeled here via the Karmarkar condition) and the isotropic HESS compact object (modeled in \cite{Panotopoulos:2024imo} via the Tolman IV solution) of very similar stellar mass and radius. Moreover, the large frequency separations are displayed in Fig. \ref{fig:3}, while the corresponding eigenfunctions $\xi(r), \eta(r)$ are shown in Fig. \ref{fig:4}. We observe that the large difference separations tend to a certain constant at highly excited modes, in agreement to the asymptotic theory \cite{tassoul}
\begin{equation}
\Delta \nu \approx \left(2 \int_0^R \frac{dr}{c_s(r)} \right)^{-1},
\end{equation}
and so a larger speed of sound implies a higher constant frequency separation.

\smallskip

\begin{figure*}[ht!]
\centering
\includegraphics[scale=1.2]{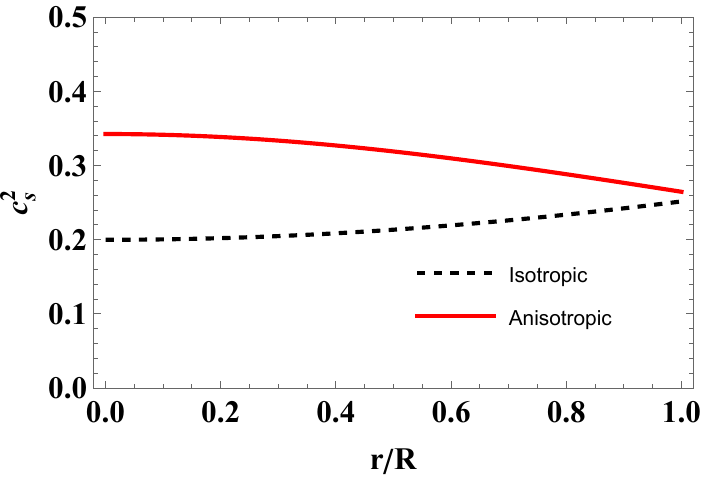} 
\caption{
Speed of sound versus radial coordinate for anisotropic star (red solid curve) and isotropic object (black dashed curve).
}
\label{fig:5} 	
\end{figure*}

The radial profiles $\xi(r),\eta(r)$ for isotropic and anisotropic stars are qualitatively very similar, the main difference being the value of the $\eta$ function at the origin, due to the fact that the relativistic adiabatic index is different from one model to another. Here we have displayed the eigenfunctions of the isotropic star. We notice that the radial profiles of the perturbations exhibit the typical behavior observed in any Sturm-Liouville problem, namely the number of zeros of the eigenfunctions equals the value of the overtone number $n=0,1,2,...$. Our findings show that the spectrum of the anisotropic object is characterized by higher frequencies, and also by larger constant frequency separation at highly excited modes, $\Delta \nu^{iso} \approx 5.36 \: kHz$ in the case of isotropic star, and $\Delta \nu^{aniso} \approx 6.48 \: kHz$ in the case of anisotropic star. Finally, for the sake of comparison, in Fig. \ref{fig:5} we have displayed the speed of sounds versus the normalized radial coordinate both for the isotropic and the anisotropic HESS compact object. The dashed curve corresponds to the isotropic star (Tolman IV solution), while the solid curve corresponds to the radial sound speed of the anisotropic object (via Karmarkar condition in gravity). Both are finite, monotonic functions of $r$ within the range $(0,1)$. In the case of isotropic star $c_s^2$ increases, whereas in the case of the anisotropic star the sound speed decreases with $r$. The speed of sound of the anisotropic object lies above the sound speed of its isotropic counterpart, which explains why $\Delta \nu^{aniso} > \Delta \nu^{iso}$.

\section{Concluding Remarks}

To summarize our work, in this article we proposed to model the HESS J1731-347 compact object (of known mass and radius) assuming that the star is made of anisotropic matter and imposing the Karmarkar condition in gravity, and we studied its radial oscillation modes. First, the general description of relativistic stars made of anisotropic matter within Einstein's gravity was reviewed. Next, we obtained an exact analytic solution via the Karmarkar condition for anisotropic stars. The three free parameters of the solution were determined imposing the appropriate matching conditions at the surface of the star of known mass and radius. Although the system of algebraic equations is non-linear, we found a unique solution. 

\smallskip

The mass-to-radius profiles as well as the $p-\rho$ relationships were shown. Then, the behavior of the solution was displayed graphically. The quantities of interest, such as relativistic adiabatic index, energy density, anisotropic factor, speed of sound etc, were found to be finite at the origin and at the same time smooth, continuous functions of the radial coordinate throughout the star. Moreover, well-established criteria, such as energy conditions, causality and stability based on the adiabatic relativistic index, were shown to be fulfilled. Since the energy density and both pressures were found to be positive, and also the energy density was higher than both radial and tangential pressure, all four energy conditions are satisfied. For comparison reasons we showed in the same figure the energy density and the pressures for the anisotropic HESS object as well as its isotropic counterpart. It is confirmed that our proposal of a Karmarkar solution to the HESS object qualitatively mimics to a strange quark star, which was observed already in \cite{ishfaq,camila}. One limitation of our Karmarkar solution, however, is that other constraints on neutron stars are not being considered in the present work. Massive radio pulsars, the NICER, the gravitational waves GW170817 and GW190425 from neutron star mergers, and their electromagnetic counterparts kilonova AT2017gfo and gamma ray bursts 170817A are being used to put constraints on neutron stars, see e.g. \cite{multi1,multi2,multi3}. Finally, given the new TOV results shown in the middle panel of Fig. \ref{fig:0}, since according to the blue curve $M_{max} \approx 3 M_{\odot}$, it can be confirmed that the obtained $M-R$ relationship is compatible with the original Drago proposal.

\smallskip

Next, regarding the spectrum of pulsating stars, we solved the equations for the perturbations imposing appropriate boundary conditions, and we computed i) the frequencies of the 8 lowest modes, ii) the corresponding eigenfunctions, and iii) the large frequency separations. Finally, for the sake of comparison we also studied the radial oscillation modes of the same object modeled via the Tolman IV solution for isotropic objects. Our findings indicate that the spectrum of the star made of anisotropic matter is characterized by higher frequencies, in agreement with the fact that its speed of sound was found to be higher than that of its isotropic counterpart.

\section{Acknowledgments}

The author wishes to thank the anonymous referees for useful comments and suggestions.



\begin{figure*}[ht!]
\centering
\includegraphics[scale=1.2]{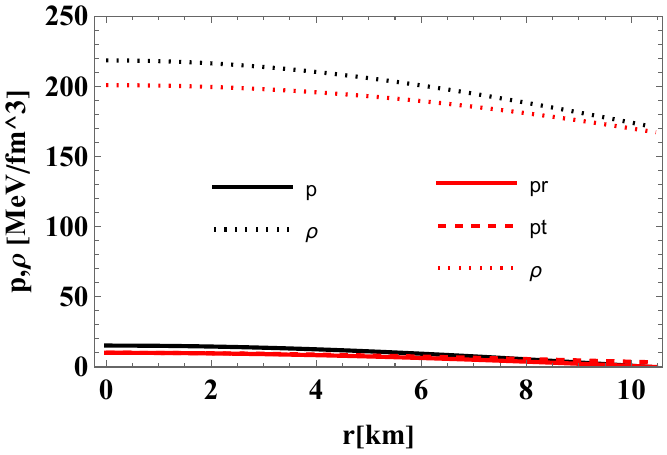} \\
\includegraphics[scale=1.2]{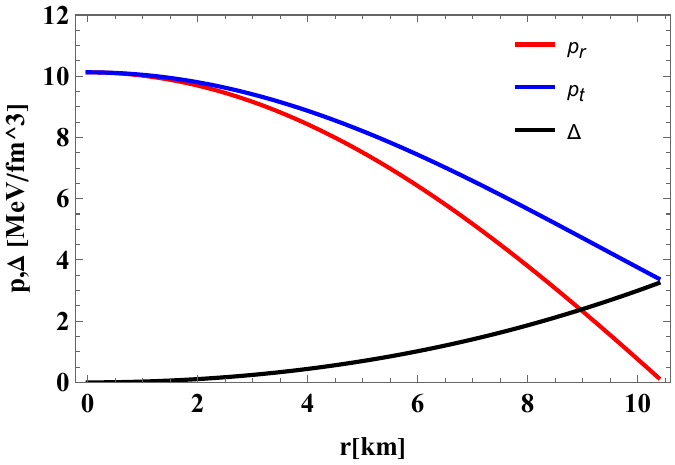} 
\caption{
Energy density and pressures (upper panel), and pressures and anisotropic factor (lower panel) versus radial coordinate in km, for the HESS compact object. The radial pressure vanishes at $r=R$, while the anisotropic factor vanishes at the origin, $\Delta(0)=0$. In the upper panel the red curves correspond to the analytic solution discussed here, whereas the black curves correspond to a numerical solution assuming the MIT bag model SQSB40 for quark matter.
}
\label{fig:1} 	
\end{figure*}


\begin{figure*}[ht!]
\centering
\includegraphics[scale=1]{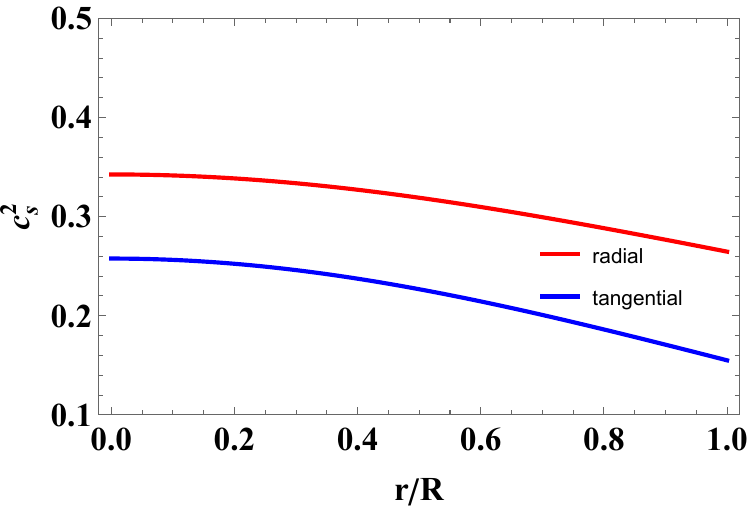} \\
\includegraphics[scale=1]{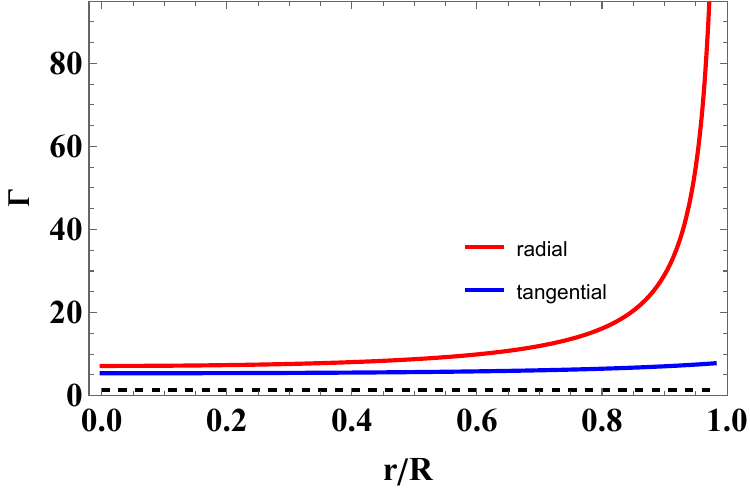}
\caption{
Speed of sound (upper panel) and relativistic adiabatic index (lower panel) versus dimensionless radial coordinate, $r/R$, for the HESS compact object. The horizontal dashed line corresponds to the critical value of $\Gamma_{cr}=1.43$. The red curves correspond to radial direction, while the blue curves correspond to tangential direction.
}
\label{fig:2} 	
\end{figure*}


\begin{thebibliography}{99}
%
\bibitem{Einstein:1915ca} A.~Einstein,
``The Field Equations of Gravitation,''
Sitzungsber. Preuss. Akad. Wiss. Berlin (Math. Phys.) \textbf{1915}, 844-847 (1915)

\bibitem{Weinberg} S. Weinberg, \textit{Gravitation and Cosmology: Principles and Applications of the General Theory of Gravitation}, John Wiley and Sons, 1972.

\bibitem{MTW} C. W. Meisner, K. S. Thorne and J. A. Wheeler, \textit{Gravitation}, Princeton University Press, 1973.

\bibitem{Ligo1} B.~P.~Abbott \textit{et al.} [LIGO Scientific and Virgo],
``Observation of Gravitational Waves from a Binary Black Hole Merger,''
Phys. Rev. Lett. \textbf{116}, no.6, 061102 (2016)
[arXiv:1602.03837 [gr-qc]].

\bibitem{Ligo2} B.~P.~Abbott \textit{et al.} [LIGO Scientific and Virgo],
``GW151226: Observation of Gravitational Waves from a 22-Solar-Mass Binary Black Hole Coalescence,''
Phys. Rev. Lett. \textbf{116}, no.24, 241103 (2016)
[arXiv:1606.04855 [gr-qc]].

\bibitem{Ligo3} B.~P.~Abbott \textit{et al.} [LIGO Scientific and VIRGO],
``GW170104: Observation of a 50-Solar-Mass Binary Black Hole Coalescence at Redshift 0.2,''
Phys. Rev. Lett. \textbf{118}, no.22, 221101 (2017)
[erratum: Phys. Rev. Lett. \textbf{121}, no.12, 129901 (2018)]
[arXiv:1706.01812 [gr-qc]].

\bibitem{Asmodelle:2017sxn} E.~Asmodelle,
``Tests of General Relativity: A Review,''
[arXiv:1705.04397 [gr-qc]].

\bibitem{ExactSol} H. Stephani, D. Kramer, M. Maccallum, C. Hoenselaers and E. Herlt, \textit{Exact Solutions of Einstein’s Field Equations},
Cambridge University Press, Cambridge, England, 2003.


\bibitem{aniso1} R. Ruderman, 
``Pulsars: Structure and dynamics,''
Ann. Rev. Astron. Astrophys., {\bf 10} (1972) 427.

\bibitem{aniso2} R. L. Bowers and E. P. T. Liang, 
``Anisotropic Spheres in General Relativity,''
Astrophys. J., {\bf 188} (1974) 657.

\bibitem{aniso3} A. I. Sokolov, 
``Phase transitions in a superfluid neutron liquid,''
JETP {\bf 79} (1980) 1137.

\bibitem{aniso4} R. F. Sawyer, 
``Condensed $\pi^-$ Phase in Neutron Star Matter,''
Phys. Rev. Lett., {\bf 29} (1972) 823.

\bibitem{aniso5} R. Kippenhahn and A. Weigert, \textit{Stellar structure and evolution}, Springer, Berlin, 1990.

\bibitem{Ovalle:2017fgl} J.~Ovalle,
``Decoupling gravitational sources in general relativity: from perfect to anisotropic fluids,''
Phys. Rev. D \textbf{95}, no.10, 104019 (2017)
[arXiv:1704.05899 [gr-qc]].

\bibitem{Ovalle:2007bn} J.~Ovalle,
``Searching exact solutions for compact stars in braneworld: A Conjecture,''
Mod. Phys. Lett. A \textbf{23}, 3247-3263 (2008)
[arXiv:gr-qc/0703095 [gr-qc]].

\bibitem{RS1} L.~Randall and R.~Sundrum,
``A Large mass hierarchy from a small extra dimension,''
Phys. Rev. Lett. \textbf{83}, 3370-3373 (1999)
[arXiv:hep-ph/9905221 [hep-ph]].

\bibitem{RS2} L.~Randall and R.~Sundrum,
``An Alternative to compactification,''
Phys. Rev. Lett. \textbf{83}, 4690-4693 (1999)
[arXiv:hep-th/9906064 [hep-th]].

\bibitem{Estrada:2018zbh} M.~Estrada and F.~Tello-Ortiz,
``A new family of analytical anisotropic solutions by gravitational decoupling,''
Eur. Phys. J. Plus \textbf{133}, no.11, 453 (2018)
[arXiv:1803.02344 [gr-qc]].

\bibitem{Morales:2018urp} E.~Morales and F.~Tello-Ortiz,
``Compact Anisotropic Models in General Relativity by Gravitational Decoupling,''
Eur. Phys. J. C \textbf{78}, no.10, 841 (2018)
[arXiv:1808.01699 [gr-qc]].

\bibitem{Estrada:2018vrl} M.~Estrada and R.~Prado,
``The Gravitational decoupling method: the higher dimensional case to find new analytic solutions,''
Eur. Phys. J. Plus \textbf{134}, no.4, 168 (2019)
[arXiv:1809.03591 [gr-qc]].

\bibitem{Ovalle:2018umz} J.~Ovalle, R.~Casadio, R.~d.~Rocha, A.~Sotomayor and Z.~Stuchlik,
``Black holes by gravitational decoupling,''
Eur. Phys. J. C \textbf{78}, no.11, 960 (2018)
[arXiv:1804.03468 [gr-qc]].

\bibitem{Ovalle:2008se} J.~Ovalle,
``Non-uniform Braneworld Stars: An Exact Solution,''
Int. J. Mod. Phys. D \textbf{18}, 837-852 (2009)
[arXiv:0809.3547 [gr-qc]].

\bibitem{Ovalle:2010zc} J.~Ovalle,
``The Schwarzschild's Braneworld Solution,''
Mod. Phys. Lett. A \textbf{25}, 3323-3334 (2010)
[arXiv:1009.3674 [gr-qc]].

\bibitem{Casadio:2012pu} R.~Casadio and J.~Ovalle,
``Brane-world stars and (microscopic) black holes,''
Phys. Lett. B \textbf{715}, 251-255 (2012)
[arXiv:1201.6145 [gr-qc]].

\bibitem{Casadio:2012rf} R.~Casadio and J.~Ovalle,
``Brane-world stars from minimal geometric deformation, and black holes,''
Gen. Rel. Grav. \textbf{46}, 1669 (2014)
[arXiv:1212.0409 [gr-qc]].

\bibitem{Ovalle:2014uwa} J.~Ovalle, L.~\'A.~Gergely and R.~Casadio,
``Brane-world stars with a solid crust and vacuum exterior,''
Class. Quant. Grav. \textbf{32}, 045015 (2015)
[arXiv:1405.0252 [gr-qc]].

\bibitem{Ovalle:2018gic} J.~Ovalle,
``Decoupling gravitational sources in general relativity: The extended case,''
Phys. Lett. B \textbf{788}, 213-218 (2019)
[arXiv:1812.03000 [gr-qc]].

\bibitem{Fernandes-Silva:2019fez} A.~Fernandes-Silva, A.~J.~Ferreira-Martins and R.~da Rocha,
``Extended quantum portrait of MGD black holes and information entropy,''
Phys. Lett. B \textbf{791}, 323-330 (2019)
[arXiv:1901.07492 [hep-th]].

\bibitem{herrera} L.~Herrera,
``New definition of complexity for self-gravitating fluid distributions: The spherically symmetric, static case,''
Phys. Rev. D \textbf{97}, no.4, 044010 (2018)
[arXiv:1801.08358 [gr-qc]].

\bibitem{comp1} G.~Abbas and H.~Nazar,
Eur. Phys. J. C \textbf{78}, no.6, 510 (2018)
[arXiv:1806.05042 [gr-qc]].

\bibitem{comp2} M.~Sharif and I.~I.~Butt,
``Complexity Factor for Charged Spherical System,''
Eur. Phys. J. C \textbf{78}, no.8, 688 (2018)
[arXiv:1808.00903 [gr-qc]].

\bibitem{comp3} G.~Abbas and H.~Nazar,
``Complexity Factor For Anisotropic Source in Non-minimal Coupling Metric $f(R)$ Gravity,''
Eur. Phys. J. C \textbf{78}, no.11, 957 (2018)
[arXiv:1811.04858 [gr-qc]].

\bibitem{comp4} H.~Nazar and G.~Abbas,
``Complexity factor for dynamical spherically symmetric fluid distributions in f(R) gravity,''
Int. J. Geom. Meth. Mod. Phys. \textbf{16}, no.11, 1950170 (2019).

\bibitem{comp5} M.~Sharif and A.~Majid,
``Complexity factor for static sphere in self-interacting Brans\textendash{}Dicke gravity,''
Chin. J. Phys. \textbf{61}, 38-46 (2019)
[arXiv:1910.06105 [gr-qc]].

\bibitem{comp6} M.~Sharif, A.~Majid and M.~M.~M.~Nasir,
``Complexity factor for self-gravitating system in modified Gauss\textendash{}Bonnet gravity,''
Int. J. Mod. Phys. A \textbf{34}, no.32, 1950210 (2019).

\bibitem{comp7} S.~Khan, S.~A.~Mardan and M.~A.~Rehman,
``Framework for generalized polytropes with complexity factor,''
Eur. Phys. J. C \textbf{79}, no.12, 1037 (2019).

\bibitem{comp8} H.~Nazar, A.~H.~Alkhaldi, G.~Abbas and M.~R.~Shahzad,
``Complexity factor for anisotropic self-gravitating sphere in Rastall gravity,''
Int. J. Mod. Phys. A \textbf{36}, no.31n32, 2150233 (2021).

\bibitem{comp9} C.~Arias, E.~Contreras, E.~Fuenmayor and A.~Ramos,
``Anisotropic star models in the context of vanishing complexity,''
Annals Phys. \textbf{436}, 168671 (2022)
[arXiv:2208.10594 [gr-qc]].

\bibitem{comp10} A.~Rincon, G.~Panotopoulos and I.~Lopes,
``Anisotropic Quark Stars with an Interacting Quark Equation of State within the Complexity Factor Formalism,''
Universe \textbf{9}, no.2, 72 (2023)
[arXiv:2301.13684 [gr-qc]].

\bibitem{comp11} \'A.~Rinc\'on, G.~Panotopoulos and I.~Lopes,
``Anisotropic stars made of exotic matter within the complexity factor formalism,''
Eur. Phys. J. C \textbf{83}, no.2, 116 (2023)
[arXiv:2302.00125 [gr-qc]].

\bibitem{karmarkar} K. R. Karmarkar, 
``Gravitational metrics of spherical symmetry and class one,''
Proc. Ind. Acad. Sci. A {\bf 27}, 56 (1948).

\bibitem{kar1} S.~K.~Maurya, S.~T.T., Y.~K.~Gupta and F.~Rahaman,
``A new exact anisotropic solution of embedding class one,''
Eur. Phys. J. A \textbf{52}, no.7, 191 (2016)
[arXiv:1512.01667 [gr-qc]].

\bibitem{kar2} K.~N.~Singh, P.~Bhar and N.~Pant,
``A new solution of embedding class I representing anisotropic fluid sphere in general relativity,''
Int. J. Mod. Phys. D \textbf{25}, no.14, 1650099 (2016)
[arXiv:1604.01013 [gr-qc]].

\bibitem{kar3} P.~Bhar, S.~K.~Maurya, Y.~K.~Gupta and T.~Manna,
``Modelling of anisotropic compact stars of embedding class one,''
Eur. Phys. J. A \textbf{52}, no.10, 312 (2016)
[arXiv:1604.00531 [gr-qc]].

\bibitem{kar4} S.~K.~Maurya, Y.~K.~Gupta, S.~Ray and D.~Deb,
``A new model for spherically symmetric charged compact stars of embedding class 1,''
Eur. Phys. J. C \textbf{77}, no.1, 45 (2017)
[arXiv:1605.01268 [gr-qc]].

\bibitem{kar5} P.~Bhar and M.~Govender,
``Anisotropic charged compact star of embedding class I,''
Int. J. Mod. Phys. D \textbf{26}, no.06, 1750053 (2016)

\bibitem{kar6} S.~K.~Maurya and S.~D.~Maharaj,
``Anisotropic fluid spheres of embedding class one using Karmarkar condition,''
Eur. Phys. J. C \textbf{77}, no.5, 328 (2017)
[arXiv:1702.04192 [physics.gen-ph]].

\bibitem{kar7} P.~Bhar, K.~N.~Singh and T.~Manna,
``A new class of relativistic model of compact stars of embedding class I,''
Int. J. Mod. Phys. D \textbf{26}, no.09, 1750090 (2017)
[arXiv:1703.03289 [gr-qc]].

\bibitem{kar8} F.~Tello-Ortiz, S.~K.~Maurya, A.~Errehymy, K.~N.~Singh and M.~Daoud,
``Anisotropic relativistic fluid spheres: an embedding class I approach,''
Eur. Phys. J. C \textbf{79}, no.11, 885 (2019).

\bibitem{kar9} M.~K.~Jasim, S.~K.~Maurya and A.~S.~M.~Al-Sawaii,
``A generalised embedding class one static solution describing anisotropic fluid sphere,''
Astrophys. Space Sci. \textbf{365}, no.1, 9 (2020).

\bibitem{kar10} L.~Baskey, S.~Das and F.~Rahaman,
``An analytical anisotropic compact stellar model of embedding class I,''
Mod. Phys. Lett. A \textbf{36}, no.05, 2150028 (2021)
[arXiv:2012.14147 [gr-qc]].

\bibitem{kar11} M.~Zubair, S.~Waheed and H.~Javaid,
``A Generic Embedding Class-I Model via Karmarkar Condition in Gravity,''
Adv. Astron. \textbf{2021}, 6685578 (2021).

\bibitem{Demorest} P.~Demorest, T.~Pennucci, S.~Ransom, M.~Roberts and J.~Hessels,
``Shapiro Delay Measurement of A Two Solar Mass Neutron Star,''
Nature \textbf{467}, 1081-1083 (2010)
[arXiv:1010.5788 [astro-ph.HE]].

\bibitem{Antoniadis} J.~Antoniadis, P.~C.~C.~Freire, N.~Wex, T.~M.~Tauris, R.~S.~Lynch, M.~H.~van Kerkwijk, M.~Kramer, C.~Bassa, V.~S.~Dhillon and T.~Driebe, \textit{et al.}
``A Massive Pulsar in a Compact Relativistic Binary,''
Science \textbf{340}, 6131 (2013)
[arXiv:1304.6875 [astro-ph.HE]].

\bibitem{recent} A.~G.~Sullivan and R.~W.~Romani,
``A Joint X-ray and Optical Study of the Massive Redback Pulsar J2215+5135,''
[arXiv:2405.13889 [astro-ph.HE]].

\bibitem{Weber} F.~Weber,
``Strange quark matter and compact stars,''
Prog. Part. Nucl. Phys. \textbf{54}, 193-288 (2005)
[arXiv:astro-ph/0407155 [astro-ph]].

\bibitem{Maxim} A.~Aziz, S.~Ray, F.~Rahaman, M.~Khlopov and B.~K.~Guha,
``Constraining values of bag constant for strange star candidates,''
Int. J. Mod. Phys. D \textbf{28}, no.13, 1941006 (2019)
[arXiv:1906.00063 [gr-qc]].

\bibitem{pulsar1} M.~C.~Miller, F.~K.~Lamb, A.~J.~Dittmann, S.~Bogdanov, Z.~Arzoumanian, K.~C.~Gendreau, S.~Guillot, W.~C.~G.~Ho, J.~M.~Lattimer and M.~Loewenstein, \textit{et al.}
``The Radius of PSR J0740+6620 from NICER and XMM-Newton Data,''
Astrophys. J. Lett. \textbf{918}, no.2, L28 (2021)
[arXiv:2105.06979 [astro-ph.HE]].

\bibitem{pulsar2} T.~E.~Riley, A.~L.~Watts, P.~S.~Ray, S.~Bogdanov, S.~Guillot, S.~M.~Morsink, A.~V.~Bilous, Z.~Arzoumanian, D.~Choudhury and J.~S.~Deneva, \textit{et al.}
``A NICER View of the Massive Pulsar PSR J0740+6620 Informed by Radio Timing and XMM-Newton Spectroscopy,''
Astrophys. J. Lett. \textbf{918}, no.2, L27 (2021)
[arXiv:2105.06980 [astro-ph.HE]].

\bibitem{pulsar3} T.~Salmi, S.~Vinciguerra, D.~Choudhury, T.~E.~Riley, A.~L.~Watts, R.~A.~Remillard, P.~S.~Ray, S.~Bogdanov, S.~Guillot and Z.~Arzoumanian, \textit{et al.}
``The Radius of PSR J0740+6620 from NICER with NICER Background Estimates,''
Astrophys. J. \textbf{941}, no.2, 150 (2022)
[arXiv:2209.12840 [astro-ph.HE]].

\bibitem{hess} Victor Doroshenko, Valery Suleimanov, Gerd P{\"u}hlhofer,
and Andrea Santangelo,
"A strangely light neutron star within a supernova remnant,"
Nature Astronomy, 6(12):1444–1451, Dec 2022.

\bibitem{violeta} V.~Sagun, E.~Giangrandi, T.~Dietrich, O.~Ivanytskyi, R.~Negreiros and C.~Provid\^encia,
``What Is the Nature of the HESS J1731-347 Compact Object?,''
Astrophys. J. \textbf{958}, no.1, 49 (2023)
[arXiv:2306.12326 [astro-ph.HE]].


\bibitem{Brillante:2014lwa} A.~Brillante and I.~N.~Mishustin,
``Radial oscillations of neutral and charged hybrid stars,''
EPL \textbf{105}, no.3, 39001 (2014)
[arXiv:1401.7915 [astro-ph.SR]].

\bibitem{Kokkotas:2000up} K.~D.~Kokkotas and J.~Ruoff,
``Radial oscillations of relativistic stars,''
Astron. Astrophys. \textbf{366}, 565 (2001)
[arXiv:gr-qc/0011093 [gr-qc]].

\bibitem{Miniutti:2002bh} G.~Miniutti, J.~A.~Pons, E.~Berti, L.~Gualtieri and V.~Ferrari,
``Non-radial oscillation modes as a probe of density discontinuities in neutron stars,''
Mon. Not. Roy. Astron. Soc. \textbf{338}, 389 (2003)
[arXiv:astro-ph/0206142 [astro-ph]].

\bibitem{Panotopoulos:2017eig} G.~Panotopoulos and I.~Lopes,
``Radial oscillations of strange quark stars admixed with condensed dark matter,''
Phys. Rev. D \textbf{96}, no.8, 083013 (2017)
[arXiv:1709.06643 [gr-qc]].

\bibitem{Passamonti:2005cz} A.~Passamonti, M.~Bruni, L.~Gualtieri, A.~Nagar and C.~F.~Sopuerta,
``Coupling of radial and axial non-radial oscillations of compact stars: Gravitational waves from first-order differential rotation,''
Phys. Rev. D \textbf{73}, 084010 (2006)
[arXiv:gr-qc/0601001 [gr-qc]].

\bibitem{Passamonti:2004je} A.~Passamonti, M.~Bruni, L.~Gualtieri and C.~F.~Sopuerta,
``Coupling of radial and non-radial oscillations of relativistic stars: Gauge-invariant formalism,''
Phys. Rev. D \textbf{71}, 024022 (2005)
[arXiv:gr-qc/0407108 [gr-qc]].

\bibitem{Savonije:2007ay} G.~J.~Savonije,
``Non-radial oscillations of the rapidly rotating Be star HD 163868,''
Astron. Astrophys. \textbf{469}, 1057 (2007)
[arXiv:0705.1755 [astro-ph]].

\bibitem{VasquezFlores:2017tkp} C.~V\'asquez Flores, Z.~B.~Hall, II and P.~Jaikumar,
``Nonradial oscillation modes of compact stars with a crust,''
Phys. Rev. C \textbf{96}, no.6, 065803 (2017)
[arXiv:1708.05985 [gr-qc]].

\bibitem{VasquezFlores:2010eq} C.~Vasquez Flores and G.~Lugones,
``Radial oscillations of color superconducting self-bound quark stars,''
Phys. Rev. D \textbf{82}, 063006 (2010)
[arXiv:1008.4882 [astro-ph.HE]].



\bibitem{Ex1} L.~M.~Franco, B.~Link and R.~I.~Epstein,
``Quaking neutron stars,''
Astrophys. J. \textbf{543}, 987 (2000)
[arXiv:astro-ph/9911105 [astro-ph]].

\bibitem{Ex2} N.~Andersson, D.~I.~Jones, K.~D.~Kokkotas and N.~Stergioulas,
``R mode runaway and rapidly rotating neutron stars,''
Astrophys. J. Lett. \textbf{534}, L75 (2000)
[arXiv:astro-ph/0002114 [astro-ph]].

\bibitem{Ex3} D.~Tsang, J.~S.~Read, T.~Hinderer, A.~L.~Piro and R.~Bondarescu,
``Resonant Shattering of Neutron Star Crusts,''
Phys. Rev. Lett. \textbf{108}, 011102 (2012)
[arXiv:1110.0467 [astro-ph.HE]].

\bibitem{Ex4} C.~Chirenti, R.~Gold and M.~C.~Miller,
``Gravitational waves from f-modes excited by the inspiral of highly eccentric neutron star binaries,''
Astrophys. J. \textbf{837}, no.1, 67 (2017)
[arXiv:1612.07097 [astro-ph.HE]].

\bibitem{corot1} E.~Michel, A.~Baglin, M.~Auvergne, C.~Catala and R.~Samadi,
``CoRoT measures solar-like oscillations and granulation in stars hotter than the Sun,''
Science \textbf{322}, 558-560 (2008)
[arXiv:0812.1267 [astro-ph]].

\bibitem{corot2} B.~Mosser, K.~Belkacem, M.~J.~Goupil, A.~Miglio, T.~Morel, C.~Barban, F.~Baudin, S.~Hekker, R.~Samadi and J.~De Ridder, \textit{et al.}
``Red-giant seismic properties analyzed with CoRoT,''
Astron. Astrophys. \textbf{517}, A22 (2010)
[arXiv:1004.0449 [astro-ph.SR]].

\bibitem{corot3} B.~Mosser, K.~Belkacem, M.~J.~Goupil, E.~Michel, Y.~Elsworth, C.~Barban, T.~Kallinger, S.~Hekker, J.~DeRidder and R.~Samadi, \textit{et al.}
``The universal red-giant oscillation pattern: an automated determination with CoRoT data,''
Astron. Astrophys. \textbf{525}, L9 (2011)
[arXiv:1011.1928 [astro-ph.SR]].

\bibitem{Kepler} A.~Bischoff-Kim and R.~H.~\O{}stensen,
``Asteroseismology of the Kepler field DBV White Dwarf - It's a hot one!,''
Astrophys. J. Lett. \textbf{742}, L16 (2011)
[arXiv:1110.2803 [astro-ph.SR]].

\bibitem{CE} M.~Evans, R.~X.~Adhikari, C.~Afle, S.~W.~Ballmer, S.~Biscoveanu, S.~Borhanian, D.~A.~Brown, Y.~Chen, R.~Eisenstein and A.~Gruson, \textit{et al.}
``A Horizon Study for Cosmic Explorer: Science, Observatories, and Community,''
[arXiv:2109.09882 [astro-ph.IM]].

\bibitem{ET} M.~Punturo, M.~Abernathy, F.~Acernese, B.~Allen, N.~Andersson, K.~Arun, F.~Barone, B.~Barr, M.~Barsuglia and M.~Beker, \textit{et al.}
``The Einstein Telescope: A third-generation gravitational wave observatory,''
Class. Quant. Grav. \textbf{27}, 194002 (2010).

\bibitem{Ovalle:2013xla} J.~Ovalle and F.~Linares,
``Tolman IV solution in the Randall-Sundrum Braneworld,''
Phys. Rev. D \textbf{88}, no.10, 104026 (2013)
[arXiv:1311.1844 [gr-qc]].

\bibitem{tolman} R.~C.~Tolman,
``Static solutions of Einstein's field equations for spheres of fluid,''
Phys. Rev. \textbf{55}, 364-373 (1939).

\bibitem{OV} J.~R.~Oppenheimer and G.~M.~Volkoff,
``On massive neutron cores,''
Phys. Rev. \textbf{55}, 374-381 (1939).

\bibitem{Schwarzschild:1916uq} K.~Schwarzschild,
``On the gravitational field of a mass point according to Einstein's theory,''
Sitzungsber. Preuss. Akad. Wiss. Berlin (Math. Phys. ) \textbf{1916}, 189-196 (1916)
[arXiv:physics/9905030 [physics]].

\bibitem{Barua:1975dxq} J.~Barua and K.~D.~Krori,
``A singularity-free solution for a charged fluid sphere in general relativity,''
J. Phys. A \textbf{8}, no.4, 508 (1975).

\bibitem{Moustakidis:2016ndw} C.~C.~Moustakidis,
``The stability of relativistic stars and the role of the adiabatic index,''
Gen. Rel. Grav. \textbf{49}, no.5, 68 (2017)
[arXiv:1612.01726 [gr-qc]].

\bibitem{Ellis} S.~W.~Hawking and G.~F.~R.~Ellis,
\textit{The Large Scale Structure of Space-Time},
Cambridge University Press, 2023.

\bibitem{Wald} R. M. Wald, \textit{General Relativity}, 
Chicago University Press, 1984.

\bibitem{Frolov} V. P. Frolov and I. D. Novikov, \textit{Black hole physics: Basic concepts and new developments}.

\bibitem{Panotopoulos:2020uvq} G.~Panotopoulos, D.~Vernieri and I.~Lopes,
``Quark stars with isotropic matter in Ho\v{r}ava gravity and Einstein\textendash{}\ae{}ther theory,''
Eur. Phys. J. C \textbf{80}, no.6, 537 (2020)
[arXiv:2006.07652 [gr-qc]].

\bibitem{Balart:2023odm} L.~Balart, G.~Panotopoulos and \'A.~Rinc\'on,
``Regular Charged Black Holes, Energy Conditions, and Quasinormal Modes,''
Fortsch. Phys. \textbf{71}, no.12, 2300075 (2023)
[arXiv:2309.01910 [gr-qc]].

\bibitem{Gondek-Rosinska:2008zmv}
D.~Gondek-Rosinska and F.~Limousin,
``The final phase of inspiral of strange quark star binaries,''
[arXiv:0801.4829 [gr-qc]].

\bibitem{Drago0} A.~Drago, A.~Lavagno and G.~Pagliara,
``Can very compact and very massive neutron stars both exist?,''
Phys. Rev. D \textbf{89}, no.4, 043014 (2014)
[arXiv:1309.7263 [nucl-th]].

\bibitem{Drago1} A.~Drago, A.~Lavagno, G.~Pagliara and D.~Pigato,
``The scenario of two families of compact stars: 1. Equations of state, mass-radius relations and binary systems,''
Eur. Phys. J. A \textbf{52}, no.2, 40 (2016)
[arXiv:1509.02131 [astro-ph.SR]].

\bibitem{Drago2} A.~Drago and G.~Pagliara,
``The scenario of two families of compact stars: 2. Transition from hadronic to quark matter and explosive phenomena,''
Eur. Phys. J. A \textbf{52}, no.2, 41 (2016)
[arXiv:1509.02134 [astro-ph.SR]].


\bibitem{Panotopoulos:2024imo} G.~Panotopoulos,
``Stellar Modeling via the Tolman IV Solution: The Cases of the Massive Pulsar J0740+6620 and the HESS J1731-347 Compact Object,''
Universe \textbf{10}, no.9, 342 (2024)
[arXiv:2408.15422 [gr-qc]].

\bibitem{pert1} G. Chanmugan, 
``Radial oscillations of zero-temperature white dwarfs and neutron stars below nuclear densities,''
ApJ, {\bf 217}, 799 (1977).

\bibitem{pert2} H. M. V{\"a}th, G. Chanmugan, 
``Radial oscillations of neutron stars and strange stars,''
Astron. Astrophys. {\bf 260}, 250 (1992).

\bibitem{Arbanil:2021ahh} J.~D.~V.~Arba\~nil and G.~Panotopoulos,
``Tidal deformability and radial oscillations of anisotropic polytropic spheres,''
Phys. Rev. D \textbf{105}, no.2, 024008 (2022)
[arXiv:2112.09729 [gr-qc]].

\bibitem{tassoul} M. Tassoul,
``Asymptotic approximations for stellar nonradial pulsations,''
ApJS, 43, 469 (1980).

\bibitem{miglio} W.~J.~Chaplin and A.~Miglio,
``Asteroseismology of Solar-Type and Red-Giant Stars,''
Ann. Rev. Astron. Astrophys. \textbf{51}, 353 (2013)
[arXiv:1303.1957 [astro-ph.SR]].

\bibitem{ilidio} D.~Capelo and I.~Lopes,
``The impact of composition choices on solar evolution: age, helio- and asteroseismology, and neutrinos,''
Mon. Not. Roy. Astron. Soc. \textbf{498}, no.2, 1992-2000 (2020)
[arXiv:2010.01686 [astro-ph.SR]].

\bibitem{ishfaq} I.~A.~Rather, G.~Panotopoulos and I.~Lopes,
``Quark models and radial oscillations: decoding the HESS J1731-347 compact object\textquoteright{}s equation of state,''
Eur. Phys. J. C \textbf{83}, no.11, 1065 (2023)
[arXiv:2307.03703 [astro-ph.HE]].

\bibitem{camila} 
C.~Sep\'ulveda and G.~Panotopoulos,
``Modeling compact objects with quark matter and dark energy: A comparative study of the radial oscillation modes of HESS J1731-347 and PSR J0740+6620,''
Chin. J. Phys. \textbf{91}, 773-783 (2024).

\bibitem{multi1}
G.~Raaijmakers, S.~K.~Greif, K.~Hebeler, T.~Hinderer, S.~Nissanke, A.~Schwenk, T.~E.~Riley, A.~L.~Watts, J.~M.~Lattimer and W.~C.~G.~Ho,
``Constraints on the Dense Matter Equation of State and Neutron Star Properties from NICER\textquoteright{}s Mass\textendash{}Radius Estimate of PSR J0740+6620 and Multimessenger Observations,''
Astrophys. J. Lett. \textbf{918}, no.2, L29 (2021)
[arXiv:2105.06981 [astro-ph.HE]].

\bibitem{multi2}
P.~T.~H.~Pang, I.~Tews, M.~W.~Coughlin, M.~Bulla, C.~Van Den Broeck and T.~Dietrich,
``Nuclear Physics Multimessenger Astrophysics Constraints on the Neutron Star Equation of State: Adding NICER\textquoteright{}s PSR J0740+6620 Measurement,''
Astrophys. J. \textbf{922}, no.1, 14 (2021)
[arXiv:2105.08688 [astro-ph.HE]].

\bibitem{multi3}
S.~Ascenzi, V.~Graber and N.~Rea,
``Neutron-star measurements in the multi-messenger Era,''
Astropart. Phys. \textbf{158}, 102935 (2024)
[arXiv:2401.14930 [astro-ph.HE]].
%
\end{thebibliography}
\end{document}